\documentclass[pdflatex,sn-mathphys-num]{sn-jnl}
\makeatletter
\def\au@and{%
  \ifnum\punctcount=2
    \unskip\space and\space
  \else
    \unskip, \space
  \fi
  \global\advance\punctcount by -1
}
\makeatother

\usepackage{graphicx}%
\usepackage{multirow}%
\usepackage{amsmath,amssymb,amsfonts}
\DeclareMathOperator{\sech}{sech}
\usepackage{amsthm}%
\usepackage{mathrsfs}%
\usepackage[title]{appendix}%
\usepackage{xcolor}%
\usepackage{textcomp}%
\usepackage{manyfoot}%
\usepackage{booktabs}%
\usepackage{algorithm}%
\usepackage{algorithmicx}%
\usepackage{algpseudocode}%
\usepackage{listings}%

\newcommand{\ii}{\mathrm{i}}

\newcommand{\Heff}{\mathbf H_{\mathrm{eff}}}

\newcommand{\m}{\mathbf m}

\theoremstyle{thmstyleone}%
\theoremstyle{thmstyletwo}%

\theoremstyle{thmstylethree}%

\begin{document}

\title[Article Title]{Enhancing magneto-optical coupling and scattering  from magnetic vortices   by  optical  polarization texturing}


\author*[1]{\fnm{Vakhtang} \sur{Jandieri}}\email{jandieri.vakhtang@nitech.ac.jp}

\author[2]{\fnm{Chenglong} \sur{Jia}}\email{cljia@lzu.edu.cn}

\author[3]{\fnm{Ramaz} \sur{Khomeriki}}\email{
ramaz.khomeriki@tsu.ge}

\author[4]{\fnm{Nicolas} \sur{Tsagareli}}\email{
ntsagareli@binghamton.edu}

\author[5]{\fnm{Qian} \sur{Li}}\email{\href{mailto:qianli_mse@mail.tsinghua.edu.cn}{qianli\_mse@mail.tsinghua.edu.cn}}

\author[6]{\fnm{Douglas H.} \sur{Werner}}\email{dhw@psu.edu}

\author[7]{\fnm{Jamal} \sur{Berakdar}}\email{Jamal.Berakdar@physik.uni-halle.de}

\affil*[1]{\orgdiv{Department of Electrical and Mechanical Engineering}, \orgname{Nagoya Institute of Technology}, \orgaddress{\street{Gokiso-cho, Showa}, \city{Nagoya}, \postcode{466-8555}, \state{Aichi}, \country{Japan}}}

\affil[2]{\orgdiv{Key Laboratory for Magnetism and Magnetic Functional Materials of Ministry of Education \& Lanzhou Center for Theoretical Physics}, \orgname{Lanzhou University}, \orgaddress{\city{Lanzhou}, \country{China}}}

\affil[3]{\orgdiv{Physics Department}, \orgname{Tbilisi State University}, \orgaddress{\street{3 Chavchavadze}, \city{Tbilisi}, \postcode{0128}, \country{Georgia}}}

\affil[4]{\orgdiv{Department of Electrical and Computer Engineering}, \orgname{Binghamton University, State University of New York}, \orgaddress{\city{Binghamton},\postcode{ 13902}, \state{NY}, \country{USA}}}

\affil[5]{\orgdiv{State Key Laboratory of New Ceramics and Fine Processing, School of Materials Science and Engineering}, \orgname{Tsinghua University}, \orgaddress{\city{Beijing}, \postcode{ 100084}, \country{China}}}

\affil[6]{\orgdiv{Department of Electrical Engineering}, \orgname{The Pennsylvania State University}, \orgaddress{\city{University Park}, \postcode{16802}, \state{PA}, \country{USA}}}

\affil[7]{\orgdiv{Institut für Physik}, \orgname{Martin-Luther Universität, Halle-Wittenberg}, \orgaddress{\city{Halle-Wittenberg}, \postcode{D-06099}, \country{Germany}}}

\abstract{
Magnetic textures offer a route toward photonic functionality in which electromagnetic fields are controlled not only by material composition and geometry but also by the internal magnetic configuration. Conversely, spatial texturing of these   fields is a qualitatively new tool to increase  the effective coupling strength to magnetic matter and to act on the the vortex topology and magnetic dynamics.   Here, we investigate the coupling and scattering of structured electromagnetic fields by a magnetic vortex, focusing on how spatially non-collinear magnetization governs polarization conversion, electromagnetic energy flow, and magnetization dynamics. We develop a spherical rigorous coupled-wave method  that incorporates the spatially dependent anisotropic magnetic permeability associated with the vortex texture. The calculations show that the off-diagonal components of the magnetic susceptibility mediate conversion into complementary polarization channels that are absent from the incident field. The generated fields exhibit pronounced spatial localization associated with the vortex magnetization profile and support a nonzero azimuthal component of electromagnetic energy flow. A comparison between linearly  and radially polarized illumination reveals an approximately one-order-of-magnitude difference in the scattered fields amplitudes due to polarization texturing, demonstrating the importance of spatial overlap between the incident field and the anisotropic magnetic response. The back-action on the magnetic vortex dynamics  is quantitatively  demonstrated  by numerical simulations for the inverse Faraday effect due to the rectification of structured effective-field excitation. It is shown that polarization textured optical fields can  drive vortex-core precession and, at sufficiently strong excitation, reverse the core polarity and chirality on nanosecond time scale. Together, these results identify magnetic non-collinearity as an active degree of freedom for electromagnetic-field control and establish incident-field structuring as a means of enhancing texture-mediated light--matter coupling.
}

\keywords{Magnetic vortex,  structured optical  fields, spherical rigorous coupled-wave method, inverse Faraday effect, chirality, ultrafast magnetism, topological structure, topological dynamics}



\maketitle

\section{Introduction}\label{sec1}

The ability to control electromagnetic fields in space, time, and polarization is central to modern photonics. Metamaterials and metasurfaces enable manipulation of light at subwavelength scales in the near and far fields \cite{Yu2011,Kildishev2013,Genevet2017}. These approaches typically rely on geometry and material composition. An additional degree of freedom is available when the internal state of a material itself influences the electromagnetic response. Magnetism is particularly attractive in this regard because the magnetic configuration can, in principle, be reconfigured without changing the physical geometry of the photonic element.

In uniformly magnetized materials, the magnetization defines a single preferred direction and its influence on electromagnetic radiation is commonly described through conventional magneto-optical effects. When the magnetization is spatially non-collinear, however, different regions of the same object present different local magnetic orientations to the electromagnetic field. Such non-collinearity can arise from nanostructuring and internal magnetic interactions \cite{Schulz2022}. The resulting spatially dependent magnetic response provides an additional route for controlling the interaction between matter and electromagnetic fields: the magnetic texture can act not merely as a quantity probed optically, but as an active part of the photonic functionality.

A particularly useful example is the magnetic vortex, a well-established state in confined ferromagnetic nanostructures. The competition between exchange and magnetostatic interactions produces a curling in-plane magnetization surrounding a small out-of-plane core \cite{Yu2012,Guslienko2006VortexCoreDynamics,Guslienko2008VortexCoreSwitching,Guslienko2010NonlinearGyrotropicDynamics,Heins2026SelfInducedFloquetMagnons,Bondarenko2024HigherOrderGyromodes,Spasojevic2025MagnetoIonicVortices,Bhattacharjee2025MagneticVortex} (see Fig.~\ref{Magn_ground}). The vortex topology is characterized by the circuity, meaning the sense of circulation of the in-plane magnetization (clockwise or counterclockwise),  polarity  which is set by the  direction of the vortex-core magnetization (up or down), and the 
vortex chirality quantifying  the handedness of the full three-dimensional vortex texture and is defined by the product of circuity and polarity.  These features can be controlled in suitably designed nanostructures, making vortices promising candidates for encoding information and realizing reconfigurable magnetic states \cite{Yu2012,Guslienko2006VortexCoreDynamics,Guslienko2008VortexCoreSwitching,Guslienko2010NonlinearGyrotropicDynamics,Heins2026SelfInducedFloquetMagnons,Bondarenko2024HigherOrderGyromodes,Spasojevic2025MagnetoIonicVortices,Bhattacharjee2025MagneticVortex}. Their pronounced spatial variation also makes them a natural platform for examining how non-collinear magnetization shapes electromagnetic scattering.

The vectorial character of a vortex is important for its optical response. An electromagnetic wave interacting with a uniformly magnetized element encounters essentially one magnetic orientation, whereas a wave interacting with a vortex samples a continuously varying configuration. The coherent superposition of these local responses can therefore generate scattered fields whose spatial and polarization structure reflects the underlying magnetic texture. Structured illumination provides a complementary control parameter: by changing the spatial distribution and polarization of the incident field, one can alter its overlap with the local anisotropic magnetic response and thereby influence the efficiency with which particular scattering channels are excited. This offers a potential means of enhancing selected light--matter coupling channels without requiring a corresponding increase in the intrinsic material response.

Recent work has explored magnetic textures as functional photonic elements and shown that spatially non-collinear magnetic configurations can generate structured electromagnetic fields with nontrivial angular momentum, optical chirality, energy-flow patterns, and magnetoelectric densities \cite{Jandieri2026}. Related studies have connected magnetic-texture winding with the orbital angular momentum of reflected light through optical spin--orbit interactions \cite{luttmann2025optical}.
Experiments investigated the role of   orbital texturing of optical fields on all-optical magnetic switching, others investigated  the magnetic vortex dynamics as signified by
magnetic helicoidal dichroism in the far field. Also, 
 optical vortex lasers were utilize for fabricating twisted ferrite microcrystal array
\cite{Kaneko2024OpticalVortexFerrite,Fanciulli2025MagneticVortex,Khalid2024PhotonicAngularMomentum}.
Modifications of the intensity profile and the polarization characteristics of radially and azimuthally polarized beams upon interacting with homogeneous magnetic materials were studied recently \cite{TAMBAG2023129649,Koksal14122024,KOKSAL2026133179}.
Coupling of optical modes to magnetic-vortex excitations in microdisks, magneto-optical control of optical-vortex beams in iron-garnet films, and strong polarization transformation in magneto-plasmonic structures containing magnetic domains have also been addressed
\cite{Yamane2022,PhysRevApplied.18.054008,PhysRevB.98.241406}. The magnetic dynamic of the vortex driven by linear-polarized or near field relied on magnetoelectric coupling  strength \cite{Yu2021NearFieldVortices,Yu2020UltrafastVortex} which is notoriously  small for realistic vortices.  The magnetic response to  polarization textured  laser vector fields has been
addressed  for atoms \cite{b3zw-mwrn,PhysRevB.99.085425}, but sofar not for magnetic vortices.
Generally, accumulated  results establish that magnetic order and electromagnetic mode structure can jointly influence optical coupling and polarization. Here, we focus on a complementary problem: how the spatially varying anisotropic (linear) magnetic response of a vortex generates complementary scattered-field components, and how their excitation depends on the spatial structure of the incident field. Motivated  by the substantially enhanced coupling in the linear response regime for polarization structured fields, we investigated the back-action of the scattered field on the vortex magnetic dynamics based on the inverse Faraday  effect \cite{vanDerZiel1965,Mills1998}, which is second order in the field.    Numerical simulations demonstrate the profound influence of the polarization  structured rectification fields on the vortex magnetic structure. 


It is also important to distinguish non-collinearity from a particular global topological classification. Magnetic skyrmions are localized spin textures characterized by a nontrivial winding of the magnetization and a finite skyrmion number \cite{Nagaosa2013,Fert2017,Bogdanov2020}; their stability and dynamics have attracted substantial interest for spintronic and information-processing applications \cite{Fert2017,Liu2021}. Vortices and skyrmions have different magnetization profiles, but both are spatially non-collinear. Since electromagnetic fields couple locally to the magnetic response, polarization conversion and structured scattering may arise from non-collinearity without requiring a nonzero skyrmion number. Magnetic vortices thus provide a useful setting in which to investigate optical effects associated with local magnetic structure separately from effects that depend on global topology.

Technically, we analyze and simulate  the coupling and scattering of electromagnetic fields in the optical regime by a magnetic vortex using a spherical rigorous coupled-wave method (RCWM) based on a spherical-wave expansion and a spatially dependent anisotropic permeability. We examine the polarization-converted field components, their spatial distributions, and the associated azimuthal electromagnetic energy flow. We compare plane-wave illumination with radially polarized structured illumination to assess how the incident-field profile affects the strength of the converted channels. Beyond the linear regime, we study the role of optical field structuring on rectification effects
and perform full-fledged  micromagnetic simulations with the rectification fields  to examine vortex-core dynamics in iron disc. 
Our central result is that the magnetic texture determines which local conversion channels are available, whereas the spatial structure of the incident field governs how efficiently those channels are excited. Considerable enhancement is achievable in the magnetic excitations via photonic field structuring. The rectified field of structured photonic pulse can drive the magnetization dynamics causing, for example, polarity switching.    

\section{Methods}\label{sec11}
The coupled equations governing the magneto-static and electromagnetic dynamics are given by:
\begin{equation}
\begin{aligned}
\frac{\partial \boldsymbol{M}(\mathbf{\tilde{r}},t)}{\partial t} = -g\mu_0\left[\boldsymbol{M}(\mathbf{\tilde{r}},t) \times \boldsymbol{H}_{\text{eff}}(\mathbf{\tilde{r}},t)\right], \\
 \nabla\times \left[\boldsymbol{H} (\mathbf{\tilde{r}},t)+ \boldsymbol{H}_{\text{ext}}(\mathbf{\tilde{r}},t)\right] = \epsilon_0  \frac{\partial \boldsymbol{E}(\mathbf{\tilde{r}},t)}{\partial t}, \\
 \nabla\times \boldsymbol{E} (\mathbf{\tilde{r}},t) = -\mu_0  \frac{\partial} {\partial t}\left[\boldsymbol{H}(\mathbf{\tilde{r}},t)+\boldsymbol{H}_{\text{ext}}(\mathbf{\tilde{r}},t)+\boldsymbol{M}(\mathbf{\tilde{r}},t)\right]
\end{aligned}
\label{eq:full}\end{equation}
where $g$ is the gyromagnetic ratio, $\boldsymbol{H}$ and $\boldsymbol{H}_{\text{ext}}$ are the demagnetization and external magnetic fields, respectively, $\boldsymbol{M}$ is the magnetization vector field, $\epsilon_0$ and $\mu_0$ are the dielectric permittivity and magnetic permeability of free space. The external magnetic field $\boldsymbol{H}_{\text{ext}}$ is related to the incident field and is defined by the boundary conditions. 

The effective magnetic field $\boldsymbol{H}_{\text{eff}}$ is obtained from the functional derivative of the free energy $F(\boldsymbol{M})$  with respect to $\boldsymbol{M}$. For the magnetic vortex dynamics an appropriate model of  $F(\boldsymbol{M})$ that includes the exchange, magnetic anisotropy, and magnetostatic contributions can be represented in the following form:
\begin{equation}
\begin{aligned}
F(\boldsymbol{M}) &= \mu_0\int_V d^3\mathbf{\tilde{r}} \left[
\frac{\ell^2}{2} (\nabla \boldsymbol{M})^2 -\gamma M_z^2 \right. \left. - \frac{1}{2} \boldsymbol{M} \cdot (\boldsymbol{H} + \boldsymbol{H}_{\text{ext}}) \right]. 
\end{aligned}
\end{equation}
The effective field is given as:    
\begin{equation}
\begin{aligned}
\boldsymbol{H}_{\text{eff}} &=-\frac{\delta F}{\delta \boldsymbol{M}}=\boldsymbol{H}_{\text{ext}} + \boldsymbol{H} + (\ell k_0)^2\left(
\frac{\partial^2}{\partial r^2} + \frac{2}{r} \frac{\partial}{\partial r} - \frac{2}{r^2}
\right) \boldsymbol{M} + 2\gamma M_z \bf{z}   \end{aligned}
\end{equation}
where $\gamma\equiv K_a/\mu_0$,    $\ell=\sqrt{2A/M_s^2\mu_0}$, $A$ is an exchange stiffness, $M_s$ is saturation magnetization, $K_a$ is the magnetic anisotropy constant, $\ell$ is a characteristic exchange length, $A_s$ is an exchange constant, $r$ denotes the radial coordinate measured from the center of the spherical vortex (the global origin of the coordinate system) to the observation point and is normalized by the free-space wavenumber $k_0$, i.e. in (3) $r$ is a non-dimensional value.

\subsection{Local magnetic susceptibility}
 In this subsection, the frequency-dependent and nonlocal magnetic susceptibility tensor of the vortex is first derived, from which the corresponding permeability tensor is subsequently obtained. Although the ground-state magnetization of a magnetic vortex is commonly described in cylindrical coordinates, the present theoretical formulation is developed directly in spherical coordinates. This choice is motivated by the spherical geometry of the electromagnetic scattering problem and the use of spherical-wave expansions within the RCWM. This provides a consistent representation of the magnetic texture and electromagnetic fields and facilitates the application of boundary conditions at the spherical interfaces, as well as the calculation of the scattered electromagnetic field. Furthermore, in the present model the magnetic vortex is assumed to be localized in the equatorial plane of the sphere, which serves as the magnetically active region in the subsequent analysis.
  
 We first linearize the magnetization with respect to small dynamic perturbations induced by the electromagnetic field: $\boldsymbol{M}(\boldsymbol{r},t) = \boldsymbol{M_0}(\boldsymbol{r}) + \boldsymbol{m}_1(\boldsymbol{r},t)$, where 
$\boldsymbol{M_0}(\boldsymbol{r})$ denotes the ground-state magnetization and $\boldsymbol{m}_1(\boldsymbol{r},t)$ is a small dynamic component. In the present model, however, the dynamic magnetization is assumed to be spatially uniform inside each spherical shell. Hence, $\boldsymbol{m}_1$ depends only on time within an individual spherical shell, i.e. the magnetic dipole has the same deviation from the ground state regardless of its location in space. The equilibrium vortex configuration is described in a cylindrical coordinate system by the following expression:
\begin{align}
  {\bf{M}}_0(\rho,\phi,\nu)= [-\tanh(\nu\rho/R)\sin(\phi), \tanh(\nu\rho/R)\cos(\phi),\sech(\nu\rho/R)]
\end{align}
where $R$ is the radius of the vortex and $0<\rho/R<1$. The parameter $\nu$ controls the spatial profile of the vortex and, in particular, determines the rate at which the magnetization changes from its out-of-plane component near the vortex core to the predominantly in-plane configuration away from the core. For small values of $\nu$, the magnetization changes only gradually with the radial distance, whereas larger values of $\nu$ produce a much sharper variation between the vortex core and its outer region.

Since the proposed formulation employs spherical waves, the ground-state magnetization is transformed into spherical coordinates. The corresponding components in the spherical coordinate system are:
\begin{equation}
\begin{gathered}
M_0^{r}(\theta,r,\nu)
=\cos(\theta)\sech\left(\frac{\nu r\sin\theta}{k_0R}\right),
\qquad
M_0^{\phi}(\theta,r,\nu)
=\tanh\left(\frac{\nu r\sin\theta}{k_0R}\right),
\\[4pt]
M_0^{\theta}(\theta,r,\nu)
=-\sin(\theta)\sech\left(\frac{\nu r\sin\theta}{k_0R}\right)
\end{gathered}
\end{equation}
where $r$ is the normalized radial coordinate measured from the center of the sphere $(0<r<1)$ and  $\theta$ is the polar angle. Since the magnetic vortex is assumed to be localized in the vicinity of the equatorial plane $\theta = \pi/2$, the ground-state magnetization in this plane reduces to: $M_0^{r}=0$, $M_0^{\phi}=\tanh(\nu r/k_0R)$, $M_0^{\theta}=-\sech(\nu r/k_0R)$. Thus, the equilibrium vortex possesses no radial magnetization component. Consequently, the magnetization lies entirely within the tangent plane of the sphere and consists only of azimuthal and polar components. This equatorial configuration forms the basis for constructing the magnetic susceptibility and permeability tensors used throughout the remainder of the paper. We note that no $\varphi$-dependence is present and $\boldsymbol{r} = (\theta,r)$. 

For time-harmonic fields of the form $e^{j \omega t}$, substituting the ground-state magnetization and the effective magnetic field $\boldsymbol{H}_{\text{eff}}$ into the linearized Landau–Lifshitz–Gilbert equation:
\begin{equation}
\begin{aligned}
\frac{\partial \boldsymbol{m_1}}{\partial t}
=-g\mu_0M_s[\boldsymbol{M_0} \times \boldsymbol{H}_{\text{eff}}]
\end{aligned}
\end{equation}
the dynamic magnetic susceptibility tensor $\chi_{i,j}$ defined through $\boldsymbol{m_1}= \boldsymbol\chi(\boldsymbol{H}_{\text{ext}} + \boldsymbol{H})$, is explicitly derived in the following form:
\begin{align}
\begin{pmatrix}
m_1^r\\
m_1^\theta\\
m_1^\phi
\end{pmatrix}
&=
\frac{1}{1+4\alpha^{2}\beta\mathcal{Q}}
\begin{pmatrix}
2\alpha^2\mathcal{Q} &
-j\alpha M_0^\phi &
j\alpha M_0^\theta \\[4pt]
j\alpha M_0^\phi &
-2\alpha^2\beta (M_0^\phi)^2 &
2\alpha^2\beta M_0^\phi M_0^\theta \\[4pt]
-j\alpha M_0^\theta &
2\alpha^2\beta M_0^\phi M_0^\theta &
-2\alpha^2\beta (M_0^\theta)^2
\end{pmatrix}
\begin{pmatrix}
H_1^r\\
H_1^\theta\\
H_1^\phi
\end{pmatrix}
\label{eq:susceptibility}
\end{align}
where $\boldsymbol{H_1}\equiv  \boldsymbol{H}_{\text{ext}} + \boldsymbol{H}$, $\mathcal{Q}(r) = (\gamma-\beta(r))(M_0^\phi)^2 - \beta(M_0^\theta)^2$, $\alpha = \frac{g\mu_0M_s}{\omega}$ and $\beta(r) = \frac{(\ell\,k_0)^2}{r^2}$.

It is important to note that in the weak-coupling regime, the magnetization-induced off-diagonal
components of the susceptibility tensor are proportional to the dimensionless magnetization–field coupling parameter $\alpha$, 
whereas the leading corrections to the diagonal components are of second
order, $\mathcal{O}(\alpha^2)$.
Consequently, the off-diagonal elements provide a first-order coupling
between different polarization components of the electromagnetic field.
This coupling can therefore lead to the generation of the new field components which are absent in the incident field. In the
weak-coupling regime, these newly generated cross-polarized components are
expected to exhibit a leading-order dependence on $\alpha$, whereas the
modification of the original components is expected to be considerably weaker.

\subsection{Spherical rigorous coupled-wave analysis}
To analyze electromagnetic scattering by localized magnetic non-collinearities, we formulate here a spherical RCWM. The spherical region is discretized into concentric onion-like shell layers, where the $l$-th thin layer $(l=1 \cdot \cdot \cdot L)$ is characterized by a  magnetic permeability matrix $\mu_{ij}(r_l) = \delta_{ij} + \chi_{ij}(r_l)$. Here, $ r_l$ denotes the radial position of the $l$-th shell layer and all material parameters are assumed to be uniform within each layer. Having derived the magnetic permeability tensor from (\ref{eq:susceptibility}), we substitute it into Maxwell's equations expressed in spherical coordinates and introduce the transformed field variables of the electric $E_r^{\alpha}$ and magnetic $H_r^{\alpha}$ fields as: $ S^{\alpha}_{\theta} = r E^{\alpha}_{\theta}, \, S^{\alpha}_{\varphi} = r \sin \theta E^{\alpha}_{\varphi}, $  $  U^{\alpha}_{\theta} = \sqrt{{\mu_0}/{\epsilon_0}}\; r  H^{\alpha}_{\theta},U^{\alpha}_{\varphi} = \sqrt{{\mu_0}/{\epsilon_0}}\;  r \sin \theta H^{\alpha}_{\varphi},$  where ${\alpha}=inc,r$ represents the incident or reflected fields.  The transformed field variables are expanded into a one-dimensional Fourier series with respect to the parameter $\kappa = \cos \theta$ ($\theta = [0:\pi]$):
\begin{equation}
\Psi(r, \theta) = \sum_{i=-N}^{N} \psi_i(r) e^{ji\pi \kappa}, \quad
\Psi = \{ S_{\theta}, S_{\varphi}, U_{\theta}, U_{\varphi} \}
\label{eq:7}
\end{equation}
and a first-order system of coupled differential equations governing the radial evolution of the Fourier amplitudes $\psi_i(r)$ is obtained:   
\begin{equation}
\frac{d}{dr}
\begin{bmatrix}
{s}_{\theta}^i \\
{s}_{\varphi}^i \\
{u}_{\theta}^i \\
{u}_{\varphi}^i
\end{bmatrix}
= j \begin{bmatrix}
    \boldsymbol{\Psi}
\end{bmatrix} =
j 
\begin{bmatrix}
\Psi_{11} & \Psi_{12} & \Psi_{13} & \Psi_{14} \\
\Psi_{21} & \Psi_{22} & \Psi_{23} & \Psi_{24} \\
\Psi_{31} & \Psi_{32} & \Psi_{33} & \Psi_{34} \\
\Psi_{41} & \Psi_{42} & \Psi_{43} & \Psi_{44}
\end{bmatrix}
\begin{bmatrix}
{s}_{\theta}^i \\
{s}_{\varphi}^i \\
{u}_{\theta}^i \\
{u}_{\varphi}^i
\end{bmatrix}.
\label{eq:8}
\end{equation}
Here $N$ denotes the truncation order of the Fourier expansion. The matrix $\boldsymbol{\Psi}$ is assembled from the Fourier coefficients of the block matrices $\Psi_{qq'}$ with the $(n,m)$ entries equal to $\Psi_{qq'}^{n-m}$. For a truncation order
$N$, each field component contains $2N+1$ Fourier harmonics, and therefore the overall system matrix has dimensions  $4(2N+1)$$\times$$4(2N+1)$. Explicit representations for the block matrices are expressed through the Toeplitz matrices of the magnetic permeability tensor and are provided in  Appendix B of our recent work \cite{Jandieri2026}.  

It is worth noting that, although the vortex magnetization is concentrated
around the equatorial plane of the sphere, this localization is represented
mathematically by a narrow normalized Gaussian function centered at the
equatorial plane. The Gaussian profile is
defined as $G_{\gamma}(\kappa)
=
\frac{1}{\sqrt{\pi}\gamma}
\exp\left(-\frac{\kappa^{2}}{\gamma^{2}}\right)$,
where $\gamma$ determines the angular width of the localized magnetic
region. In the limit as $\gamma\rightarrow 0$, $G_{\gamma}(\kappa)$ approaches
the Dirac delta distribution $\delta(\kappa)$.
The localization is introduced into the permeability tensor by retaining
the free-space background in the diagonal components,
$\mu_{ii}(r,\kappa)
=
1+
\left[\mu_{ii}(r)-1\right]G_{\gamma}(\kappa),
$
whereas the off-diagonal components are written as
$\mu_{ij}(r,\kappa)
=
\mu_{ij}(r)G_{\gamma}(\kappa),
\quad i\neq j$. Consequently, away from the equatorial magnetic layer,
$G_{\gamma}(\kappa)\rightarrow 0$, hence, the permeability tensor approaches
that of free space $\mu_{ii}\rightarrow 1$ and
$\mu_{ij}\rightarrow 0 \quad (i\neq j)$.
The corresponding Fourier--Toeplitz matrix elements entering the RCWM
formulation are then obtained as:
\begin{equation}
\left[\!\left[\mu_{ij}\right]\!\right]_{pm}
=
\frac{1}{2}
\int_{-1}^{1}
\mu_{ij}(r,\kappa)
\exp\left[-{j}\pi(p-m)\kappa\right]
\,d\kappa,
\qquad
p,m=-N,\ldots,N.
\end{equation}

\begin{figure}[t]
\includegraphics[width=0.6\linewidth]{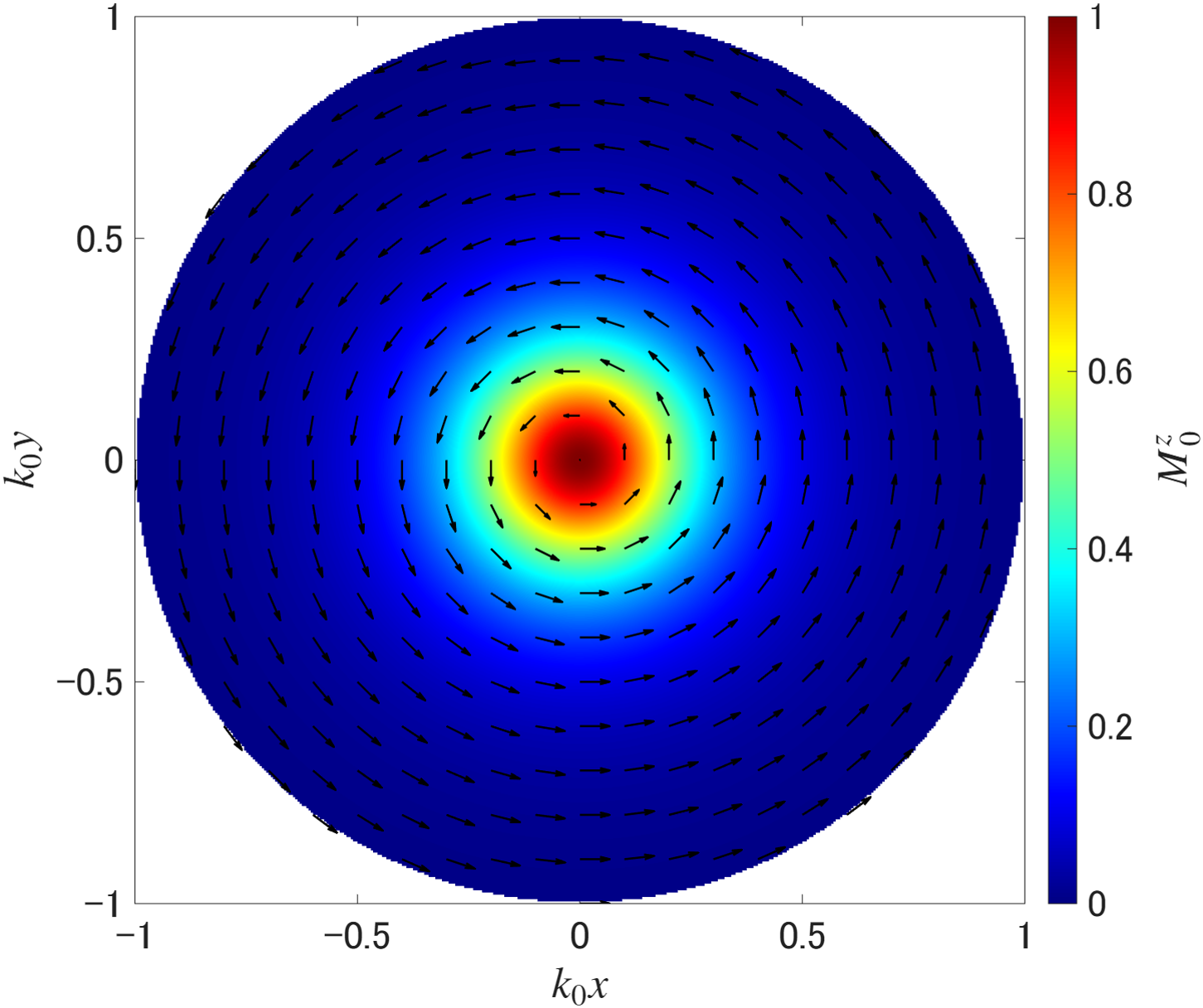} 
\caption{{ Magnetization distribution of the vortex ground state. The arrows indicate the in-plane magnetization $M_0^\phi$ in the equatorial $xoy$-plane, showing the circulating vortex structure, while the color map represents the out-of-plane component $M_0^z$. The $M_0^z$ component is maximal at the vortex center and rapidly decreases away from the core. }     }\label{Magn_ground}
\end{figure}
The Gaussian profile therefore serves as a regularized numerical approximation to
the Dirac delta distribution while providing a smooth function suitable
for the Fourier expansion employed in the RCWM formulation. Its
normalization ensures unit integrated weight in the narrow-Gaussian limit
(and effectively unit weight over $-1\leq\kappa\leq1$ for sufficiently
small $\sigma$), thereby preserving the integrated strength of the
localized magnetic response. Consequently, the electromagnetic interaction
associated with the vortex magnetization is effectively confined to a thin
equatorial magnetic layer embedded within the spherical scattering
geometry.

The solution of (\ref{eq:8}) in each anisotropic shell layer is obtained through the eigenvalues and eigenvectors of the matrix $\boldsymbol{\Psi}$. Consequently, the tangential electric $({\boldsymbol{S}_\theta}^{(l)}({\kappa,r_l}), {\boldsymbol{S}_\varphi}^{(l)}({\kappa,r_l}))$ and magnetic $({\boldsymbol{U}_\theta}^{(l)}({\kappa,r_l}), {\boldsymbol{U}_\varphi}^{(l)}({\kappa,r_l}))$  fields in the $l$-th shell can be written in the following form: 
\begin{align}
{\textbf{S}}^{(l)}({\kappa}, r_l)
 &= \sum_{i=-N}^{N}
    \Bigg(
        \sum_{p=1}^{4(2N+1)}
        P_{ip,S}^{(l)}\, e^{jr_l |\xi^l_p|}\, C_p^{(l)}
    \Bigg)
    e^{ji\pi\kappa} =
     \sum_{i=-N}^{N}
    \Big(
        \overline{\overline{\boldsymbol{P}}}_S^{(l)} \cdot
        \boldsymbol{\Upsilon} \cdot
        \boldsymbol{C}^{(l)}
    \Big)
    e^{ji\pi\kappa}
\label{eq::9}
\end{align}

\begin{align}
{\textbf{U}}^{(l)}({\kappa}, r_l)
 &= \sum_{i=-N}^{N}
    \Bigg(
        \sum_{p=1}^{4(2N+1)}
        P_{ip,U}^{(l)}\, e^{jr_l |\xi^l_p|}\, C_p^{(l)}
    \Bigg)
    e^{ji\pi\kappa} =
    \sum_{i=-N}^{N}
    \Big(
        \overline{\overline{\boldsymbol{P}}}_U^{(l)} \cdot
        \boldsymbol{\Upsilon} \cdot
        \boldsymbol{C}^{(l)}
    \Big)
    e^{ji\pi\kappa}
\label{eq::10}
\end{align}
where we introduced the shorthand notation 
\begin{align}
\boldsymbol{S}^{(l)}(\boldsymbol{\kappa},r_l)
&=
\begin{bmatrix}
\boldsymbol{S}_{\theta}^{(l)}(\boldsymbol{\kappa},r_l),
\boldsymbol{S}_{\varphi}^{(l)}(\boldsymbol{\kappa},r_l)
\end{bmatrix}^{T},\\
\boldsymbol{U}^{(l)}(\boldsymbol{\kappa},r_l)
&=
\begin{bmatrix}
\boldsymbol{U}_{\theta}^{(l)}(\boldsymbol{\kappa},r_l),
\boldsymbol{U}_{\varphi}^{(l)}(\boldsymbol{\kappa},r_l)
\end{bmatrix}^{T}.
\end{align}
The matrices
$\overline{\overline{\boldsymbol{P}}}_S^{(l)}$
and
$\overline{\overline{\boldsymbol{P}}}_U^{(l)}$ each of size $2(2N+1)$$\times$$4(2N+1)$ are constructed from the eigenvectors of the system matrix $\boldsymbol{\Psi}^l$ associated with the $l$-th thin shell. The diagonal matrix $\boldsymbol{\Upsilon}$ contains the propagation factors determined by the corresponding eigenvalues $\{\xi^l_p\}$, while the vector $\boldsymbol{C}^{(l)}$ represents the unknown modal amplitudes in the $l$-th shell. The total electromagnetic fields outside the spherical region containing the vortex are expressed as a superposition of the incident field, represented using Schelkunoff–Bessel functions, and the scattered field, represented using outgoing Schelkunoff–Hankel functions \cite{Jandieri2026}. 

At each interface between adjacent spherical shells, the electromagnetic
boundary conditions are imposed by requiring continuity of the tangential
electric- and magnetic-field components. Accordingly, the state variables
$S_{\theta}$, $S_{\phi}$, $U_{\theta}$, and $U_{\phi}$ are matched across
every internal shell boundary. At the innermost interface, in order to avoid a singularity at $r=0$, the RCWM solution
is matched to the regular solution expressed in terms of
Schelkunoff--Bessel functions, thereby enforcing regularity toward the
origin. At the outer surface of the vortex-containing region, the internal
RCWM solution is matched to the superposition of the prescribed incident
field and the outgoing scattered field represented by
Schelkunoff--Hankel functions. These boundary and interface conditions
determine the unknown modal amplitudes in all spherical shells together
with the coefficients of the regular inner and outgoing scattered fields.

\begin{figure}[t]
\includegraphics[width=0.6\linewidth]{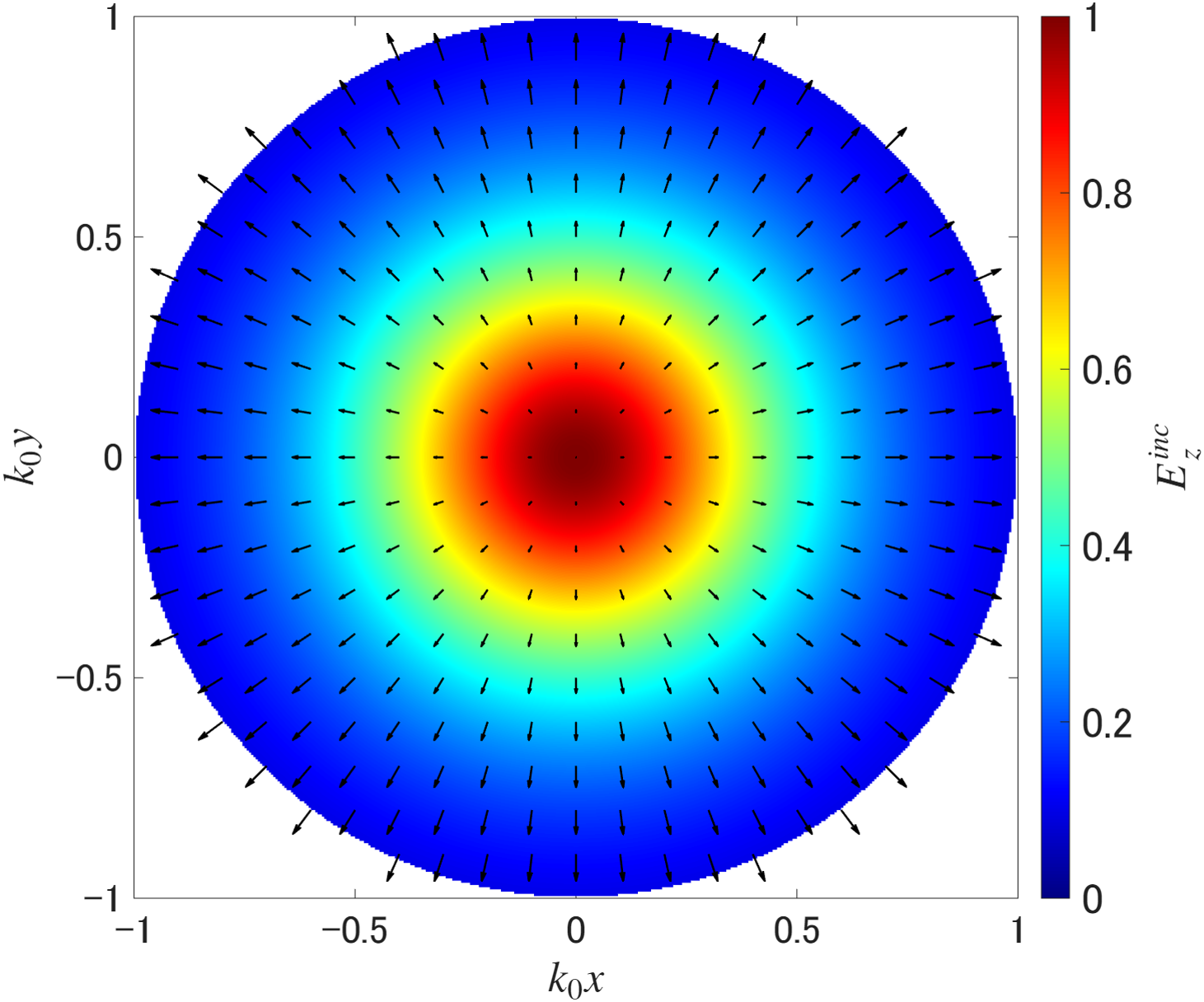} 
\caption{{Incident electric field distribution at $\sigma = 45^0$. The arrows indicate the in-plane electric-field component $E_\rho$ in the equatorial $xoy$-plane, showing the radially polarized field distribution, while the color map represents the out-of-plane component $E_z$. The $E_z$ component is maximal at the center and decreases away from the center. }  }\label{electric_field_inc}
\end{figure}

\begin{figure}[t]
\includegraphics[width=0.6\linewidth]{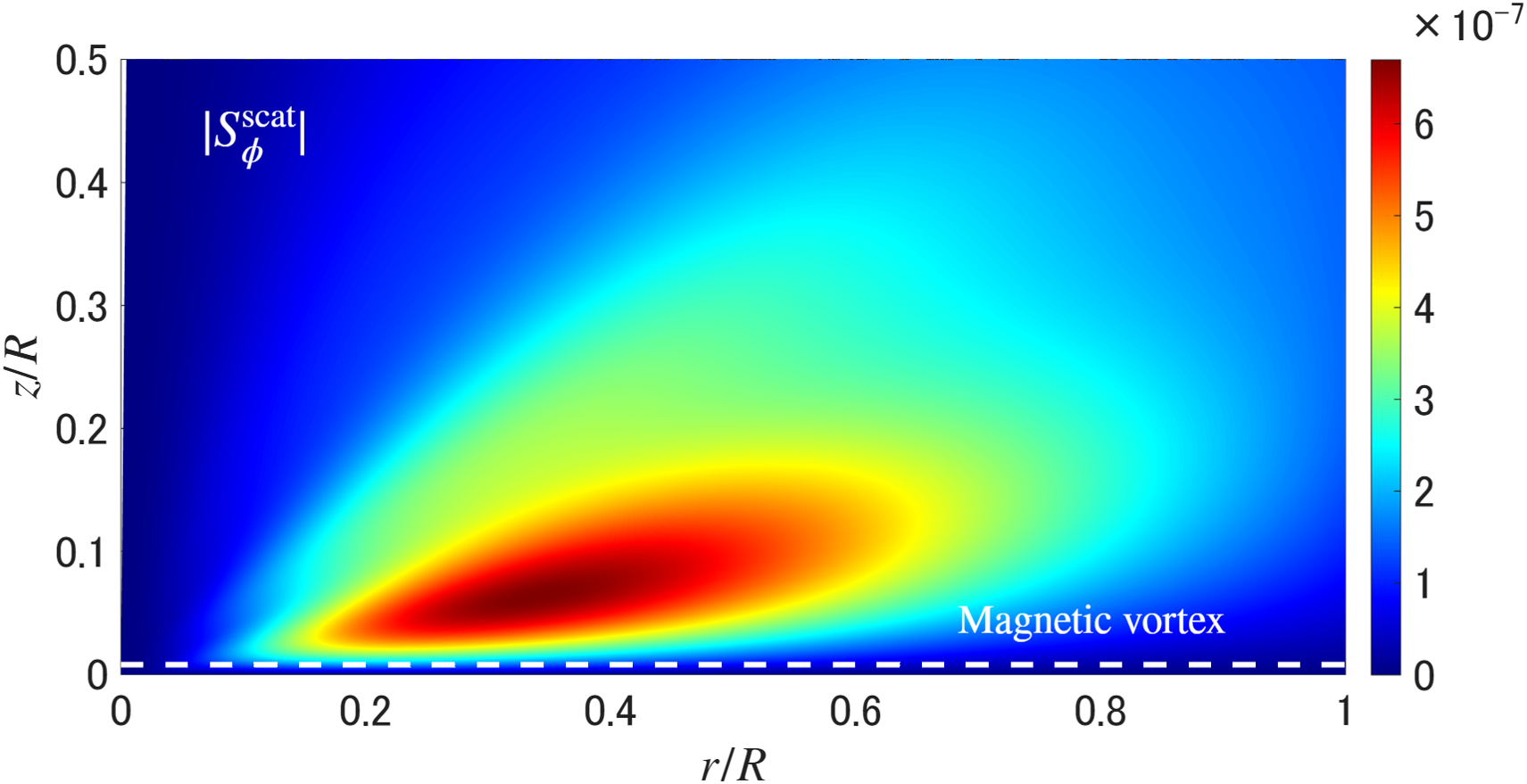} 
\caption{{Spatial distribution of the magnitude of the azimuthal component of the scattered field, $|S_{\phi}^{\mathrm{scat}}|$. The white dashed line indicates the position of the magnetic vortex in the vicinity of the equatorial plane $z/R=0$. }  }\label{electric_field_scat}
\end{figure}

\begin{figure}[t]
\includegraphics[width=0.6\linewidth]{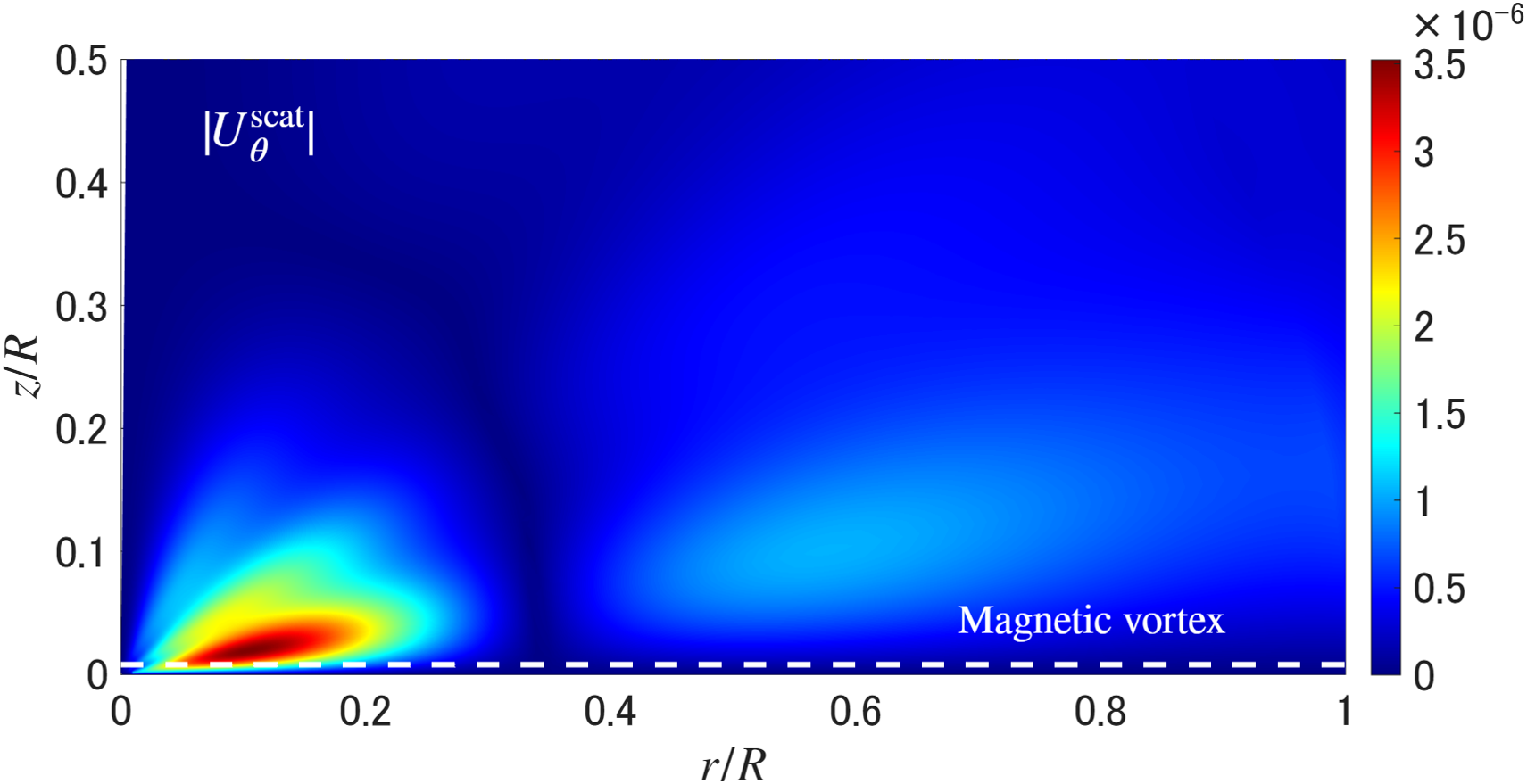} 
\caption{{Spatial distribution of the magnitude of the polar component of the scattered field, $|U_{\theta}^{\mathrm{scat}}|$. The white dashed line indicates the position of the magnetic vortex in the vicinity of the equatorial plane $z/R=0$. }  }\label{magnetic_field_scat}
\end{figure}

\begin{figure}[t]
\includegraphics[width=0.6\linewidth]{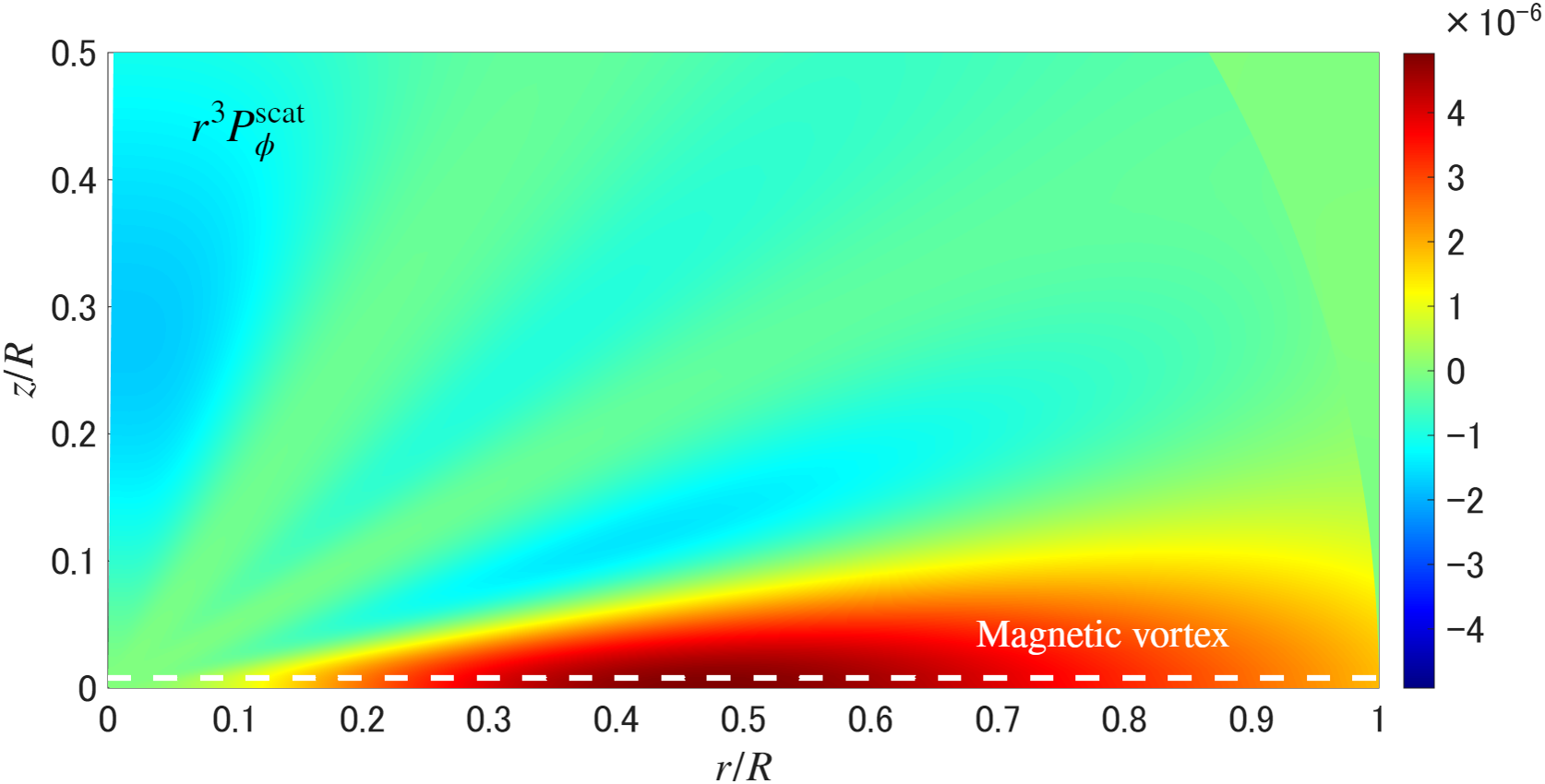} 
\caption{{Spatial distribution of the rescaled azimuthal component of the Poynting vector, $r^3P_{\phi}^{\mathrm{scat}}$. The white dashed line indicates the position of the magnetic vortex in the vicinity of the equatorial plane $z/R=0$.}  }\label{Poynting_scat_phi}
\end{figure}

\begin{figure}[t]
\includegraphics[width=0.6\linewidth]{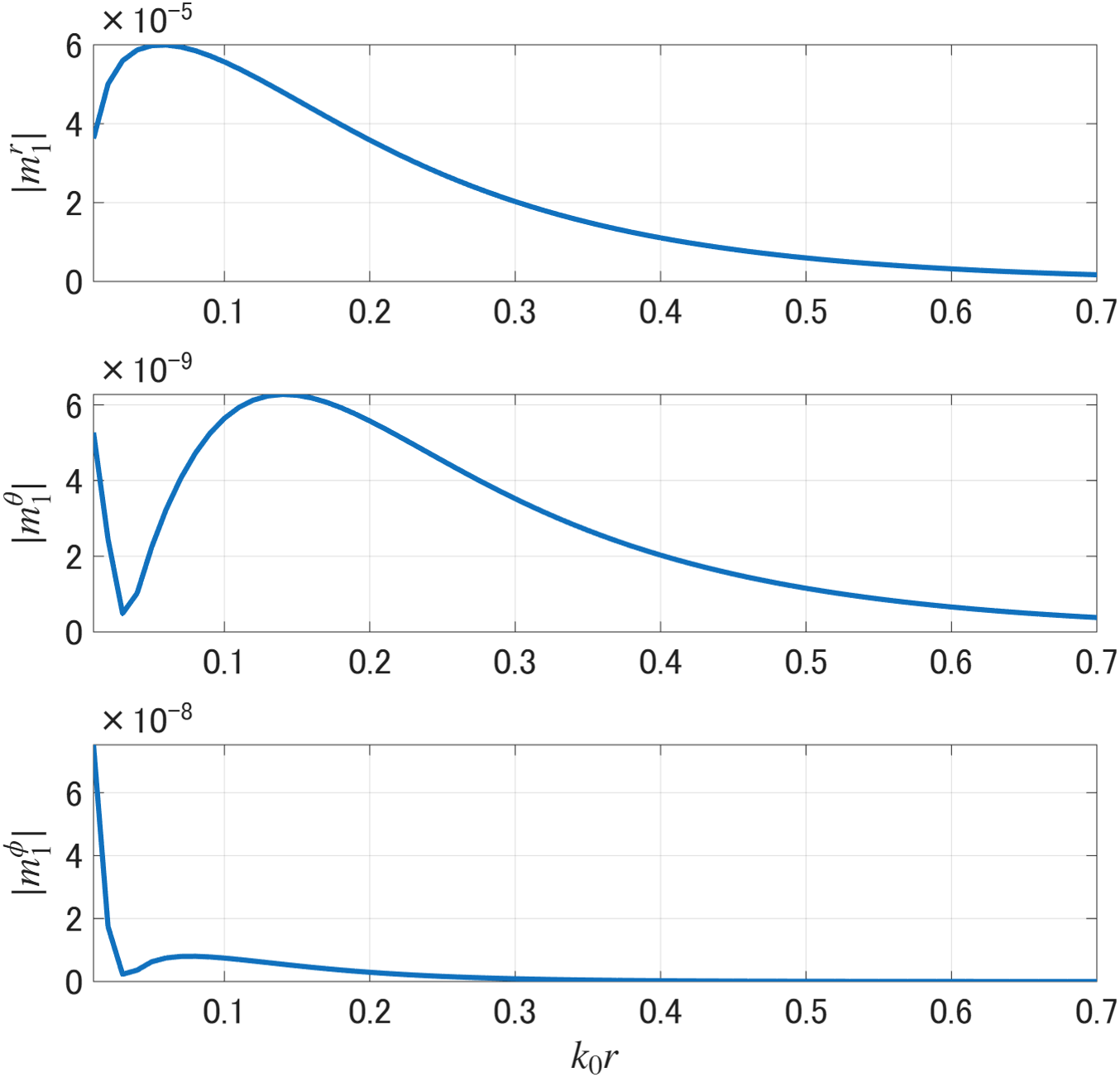} 
\caption{Radial dependence of the magnitudes of the dynamic magnetization
components $|m_1^r|$, $|m_1^\theta|$, and $|m_1^\phi|$ induced by the
radially polarized incident field at a fixed polar angle in the equatorial
plane. The dynamic response is predominantly localized toward the vortex core and decreases with increasing dimensionless radial coordinate $k_0r$.} \label{dynamical}
\end{figure}

\subsection{Incident field}
The incident electromagnetic field is taken to be a structured radially
polarized beam propagating along the $z$ direction. In cylindrical
coordinates $(\rho,\phi,z)$, the electric field possesses radial and
longitudinal components, whereas the magnetic field is purely azimuthal.
Introducing the transverse and longitudinal wave-vector components
$k_{\perp}=k_0\sin\sigma$ and $k_z=k_0\cos\sigma$, where $\sigma$
determines the angular structure of the incident beam, the nonvanishing
electric- and magnetic-field components are written as:

\begin{equation}
E_{\rho}^{\mathrm{inc}}(\rho,z)
=
J_1(k_{\perp}\rho)
\exp\!\left(jk_z z\right)
\end{equation}
\begin{equation}
E_{z}^{\mathrm{inc}}(\rho,z)
=
j\frac{k_{\perp}}{k_z}
J_0(k_{\perp}\rho)
\exp\!\left(jk_z z\right)
\end{equation}
and 
\begin{equation}
H_{\phi}^{\mathrm{inc}}(\rho,z)
=
-\frac{1}{\eta_0 cos\sigma}
J_1(k_{\perp}\rho)
\exp\!\left(jk_z z\right),
\end{equation}
where $J_0$ and $J_1$ denote Bessel
functions of the first kind of orders zero and one, respectively. The remaining cylindrical components satisfy
$E_{\phi}^{\mathrm{inc}}=H_{\rho}^{\mathrm{inc}}
=H_{z}^{\mathrm{inc}}=0$.

After transforming the incident radially polarized beam from cylindrical
coordinates $(\rho,\phi,z)$ to spherical coordinates $(r,\theta,\phi)$,
using the spherical-wave expansion in the RCWM formalism, the incident $S_\theta$ field can be written in terms of the Schelkunoff (Riccati--Bessel) functions as follows:
\begin{equation}
\begin{aligned}
S_\theta^{\mathrm{inc}}(r,\theta)
={}&
\cos\theta
\sum_{n=0}^{\infty}
{j}^{\,n}
(2n+3)
\frac{n!}{(n+2)!}\,
P_{n+1}^{1}(\cos\theta)
P_{n+1}^{1}(\cos\sigma)\,
\widehat{j}_{n+1}(r)
\\[2mm]
&\quad
-{j}\sin\theta\,\frac{k_\perp}{k_z}
\sum_{n=0}^{\infty}
{j}^{\,n}
(2n+1)\,
P_n(\cos\theta)
P_n(\cos\sigma)\,
\widehat{j}_{n}(r).
\end{aligned}
\end{equation}
The corresponding normalized azimuthal magnetic-field component is given as:
\begin{equation}
U_\phi^{\mathrm{inc}}(r,\theta)
=
-\frac{\sin\theta}{\cos\sigma}\
\sum_{n=0}^{\infty}
{j}^{\,n}
(2n+3)
\frac{n!}{(n+2)!}\,
P_{n+1}^{1}(\cos\theta)
P_{n+1}^{1}(\cos\sigma)\,
\widehat{j}_{n+1}(r).
\end{equation}
Additionally, we have a nonzero $E_r$ component of the electric field, but it does not play a role in the boundary conditions. Here $P_n$ and $P_n^1$ denote the Legendre and associated Legendre
functions, respectively, and $\widehat{j}_n$ denotes the Schelkunoff
(Riccati--Bessel) function. In the numerical implementation, the infinite sums are truncated at a finite order $N$, while the
remaining angular dependence is subsequently expanded in the Fourier
basis $\exp({j}\pi m\kappa)$ (cf. a convergence test in the Supp.\ Mat.). 

As demonstrated in the following section, the emergence of the complementary tangential components, $E_\phi$ ($S_\phi$) and $H_\theta$ ($U_\theta$) not present in the incident wave, upon interaction with the magnetized vortex provides a direct signature of polarization conversion induced by the anisotropic off-diagonal components of the magnetic permeability tensor.

\section{Results}\label{sec2}

The numerical calculations reveal a pronounced spatial correlation between
the magnetic-vortex structure and the electromagnetic response of the
system. In the numerical analysis we use the following parameters: $\alpha = 5 \cdot 10^{-5}$, $\gamma = 10^{-4}$, $\ell k_0 = 0.04$, $k_0 R = 1$ and $\nu=6$. Figure~\ref{Magn_ground} shows the ground-state magnetization in the
equatorial $xoy$ plane. In the central region of the vortex, the magnetization
possesses a strong out-of-plane component (cf. the color map), whereas with increasing radial
distance the magnetization continuously rotates toward the azimuthal
direction. In terms of the spherical components employed in the present
formulation, this behavior is described by the transition from a dominant
$M_{\theta}^{0}$ component close to the vortex core to a dominant
$M_{\phi}^{0}$ component away from the core. The spatial structure of the
incident radially polarized field is shown in
Fig.~\ref{electric_field_inc}. Although the transverse component of the
incident field (marked by arrows) becomes small toward the beam axis, the electromagnetic
interaction is not determined by the incident-field amplitude alone. The
local response is governed by the product of the electromagnetic field and
the spatially dependent anisotropic susceptibility of the vortex given by  (\ref{eq:susceptibility}). In
particular, the off-diagonal coupling terms proportional to
$\alpha M_{\theta}^{0}$ become important in the core region, where
$|M_{\theta}^{0}|$ is largest. Consequently, a relatively weak local
incident field can still produce a pronounced response when its polarization has a favorable overlap with
the locally enhanced coupling channels of the vortex.

A particularly clear manifestation of this interaction is the generation
of the complementary tangential field components
$S_{\phi}^{\mathrm{scat}}$ and $U_{\theta}^{\mathrm{scat}}$, shown in
Figs.~\ref{electric_field_scat} and \ref{magnetic_field_scat},
respectively. These components are absent in the incident radially polarized
field and therefore, arise entirely from the anisotropic magnetic response of
the vortex. Their appearance can consequently be interpreted as a
polarization-conversion effect induced by the off-diagonal elements of the
permeability tensor. The generated fields are strongly nonuniform and are
predominantly concentrated close to the equatorial magnetic layer and toward
the vortex-core region. In particular, $|U_{\theta}^{\mathrm{scat}}|$
exhibits a pronounced maximum close to the region in which the out-of-plane
ground-state magnetization is appreciable. This behavior is consistent with
the structure of the coupling matrix, since the incident $U_{\phi}$ channel
can couple to the complementary $U_{\theta}$ channel through permeability
elements containing $M_{\theta}^{0}$. As $M_{\theta}^{0}$ decreases away
from the core and $M_{\phi}^{0}$ becomes dominant, this particular
conversion channel becomes weaker. The numerical results therefore indicate
that the spatial distribution of the converted electromagnetic field is
controlled not simply by the intensity distribution of the incident beam,
but by the combined spatial overlap of the incident field with the
orientation-dependent magnetic susceptibility of the vortex. 

{For comparison, we also consider the homogeneous wave. To ensure a consistent comparison between different incident-field configurations, the radially polarized beams are normalized to carry equal incident power through a circular aperture of radius \(R\), corresponding to the transverse cross-section of the vortex-containing spherical region. The incident power is calculated by integrating the axial component of the time-averaged Poynting vector over this area, and the resulting normalization factor is applied to both the electric and magnetic incident-field amplitudes. This normalization enables a meaningful comparison between the homogeneous and strongly structured illumination regimes, allowing the influence of the incident-field spatial structure on vortex-induced scattering to be investigated independently of differences in incident power.
%
%
%
The spatial distributions of the scattered field components $|S_{\phi}^{\mathrm{scat}}|$ and $|U_{\theta}^{\mathrm{scat}}|$ exhibit profiles very similar to those obtained for the radially polarized incident wave. However, their magnitudes are much smaller (approximately one orders of magnitude) than those obtained for the structured incident field. The similar spatial distributions of the scattered components obtained for homogeneous and radially polarized illumination arise because, in both cases, the incident field excites the same nonvanishing RCWT channels, $S_{\theta}^{\mathrm{inc}}$ and $U_{\phi}^{\mathrm{inc}}$, while the complementary components $S_{\phi}^{\mathrm{scat}}$ and $U_{\theta}^{\mathrm{scat}}$ are generated through the same anisotropic off-diagonal elements of the vortex permeability tensor. 
}

\begin{figure}[t]
\includegraphics[width=0.95\linewidth]{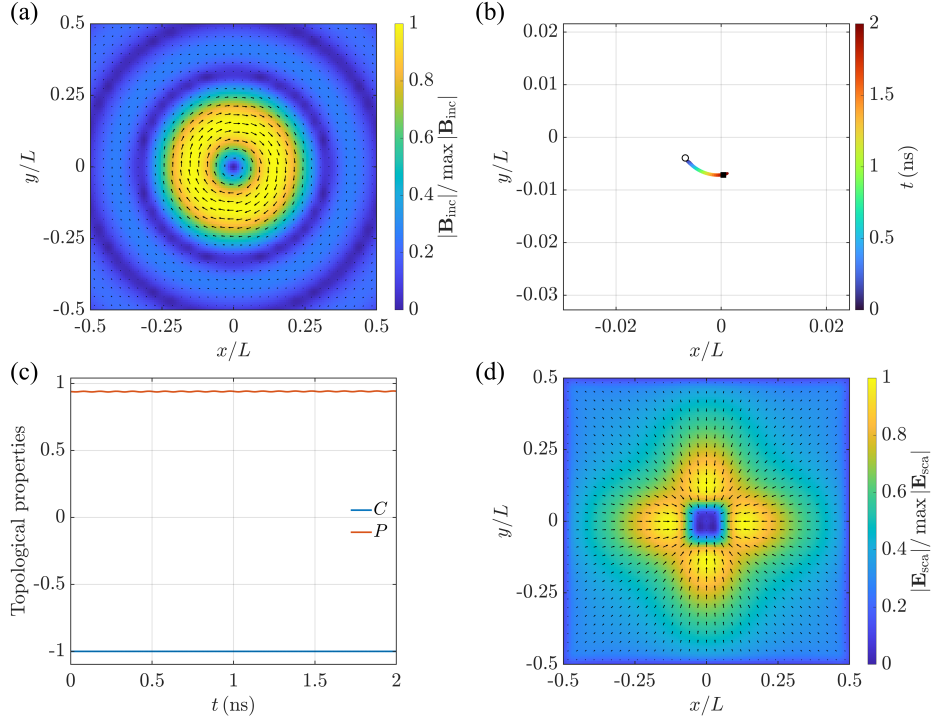} 
\caption{Micromagnetic simulations of magnetic vortex dynamics and
electromagnetic-field scattering. (a) Spatial distribution of the effective magnetic field with an amplitude of 0.1 T, generated by the interference
of two structured optical electric fields. (b) Trajectory of the magnetic vortex core under the structured-field excitation. (c) Time evolution of the vortex-core polarity ($P$) and circuity ($C$) under excitation by the effective
magnetic field shown in (a). The chirality is $\chi=P\, C$. (d) Snapshots of the spatial distribution of the emergent electric field induced by the driven vortex dynamics.
} \label{simulations}
\end{figure}

The polarization conversion is also reflected in the electromagnetic
energy flow. Figure~\ref{Poynting_scat_phi} shows the radially rescaled
azimuthal Poynting component $r^{3}P_{\phi}^{\mathrm{scat}}$, where
$P_{\phi}
=
\frac{1}{2}\operatorname{Re}
\left(
E_r H_{\theta}^{*}
-
E_{\theta}H_r^{*}
\right).$
The factor $r^{3}$ is introduced only as a visualization rescaling that
compensates for the explicit geometrical radial factors associated with the
scaled $S$ and $U$ variables. The nonzero value of $P_{\phi}$ demonstrates that the
vortex-induced scattered field possesses an azimuthal component of the
local electromagnetic energy flow. The strongest azimuthal response occurs
close to the equatorial vortex region, where the magneto-optical coupling is
most effective. The electromagnetic response can be directly related to the
dynamical perturbation of the vortex magnetization. Using the calculated reflected 
magnetic-field components $H_r$, $H_{\theta}$, and $H_{\phi}$ at a fixed
polar angle in the vicinity of the equatorial plane, the dynamic magnetization
$\mathbf{m}_1=(m_1^r,m_1^\theta,m_1^\phi)^{T}$ is obtained from the local
susceptibility relation derived above. The resulting radial profiles are
shown in Fig.~\ref{dynamical}. A clear picture between the three
components is observed, with $|m_1^r|$ providing the dominant dynamical
response over most of the relevant radial interval, whereas
$|m_1^\theta|$ and $|m_1^\phi|$ are substantially smaller. The strong
response toward the vortex core can be understood from the structure of the
susceptibility matrix. In particular, the radial component contains the
first-order coupling term $m_1^r
\propto
\frac{j\alpha M_{\theta}^{0}H_{\phi}}
     {1+4\alpha^{2}\beta Q}$, which directly couples the dominant azimuthal magnetic-field component to
the radial magnetization dynamics. Since
$|M_{\theta}^{0}|$ approaches its maximum value toward the vortex core,
this coupling channel is particularly effective there, even though the
incident radially polarized field itself becomes weaker toward the beam
axis. Away from the core, $M_{\theta}^{0}$ decreases while
$M_{\phi}^{0}$ becomes dominant; however, the corresponding first-order
coupling involves $H_{\theta}$, which is absent from the incident field and
is generated only through the scattering process. The resulting response is
therefore considerably weaker. The maximum dynamic response consequently
occurs at a small but finite radial distance, reflecting the competition
between the decreasing incident-field amplitude toward the origin and the
increasing magnetic coupling associated with the vortex core. Taken
together, these results indicate that the magnetic vortex acts not merely
as a spatially localized perturbation of the permeability, but as a
polarization-selective scattering medium that converts the incident field,
generates an azimuthal redistribution of electromagnetic energy, and
produces a localized dynamical magnetization response.
\\
\\
\\
{\textbf{Micromagnetic Simulations} \\ 
The RCWM formulation considers an idealized spherical geometry in which the anisotropic magnetic response is localized near the equatorial plane, effectively representing a thin magnetic layer.   For  a comparison with a specific real sample, we perform   full numerical  micromagnetic simulations for a square magnetic iron film. Both RCWM and numerical  descriptions are found to capture the essential non-collinear magnetization texture and linear response,  establishing a qualitative consistency of the analytical study.
A key motivation for conducting full numerical study is the ability to go beyond the linear magnetic response regime (linear in the incident field amplitude). Having established the enhanced magneto-optical coupling via optical field texturing, it is natural to wonder about the consequences for the back-action of this texturing on the magnetic vortex itself.  To this end, we study  the regime of the inverse Faraday effect. A linear resonant coupling of magnetic vortices to structured fields such as polarization structured or electromagnetic vortices
\cite{jiang2021electromagnetic} is expected to be very weak, for a diffraction-limited impinging GHz field acts on the length scale of the magnetic vortex as an almost homogeneous field.
 In the Supp.\ Mat.,  we outline how 
 the superposition of two  polarization-textured optical fields with  different frequencies can generate, through the inverse Faraday effect, an effective textured  magnetic field oscillating at their  frequency difference. When this frequency approaches the vortex excitations modes, strongly linear magnetic dynamics can be driven that is found to be  qualitatively different or even absent for linearly polarized optical pulses.\\
%
To evaluate the experimental feasibility,  we performed full three-dimensional micromagnetic simulations of a thin magnetic film hosting a vortex texture \cite{Jia2020}. The material parameters for Fe were chosen: a saturation magnetization of $M_s=1.7\times10^5$ A/m, an exchange stiffness of $A_{\mathrm{ex}}=1\times10^{-11}$ J/m, and an out-of-plane uniaxial anisotropy of $K_u=4.7\times10^4$ J/m$^3$. The Gilbert damping constant of $\alpha=0.1$ was used for fast magnetization relaxation. The simulated structure consists of a Fe square with a width of 100 nm and a thickness of 2 nm, discretized into cubic cells of size $(2~\mathrm{nm})^3$. An initial vortex configuration with positive core polarity ($P=+1$) and counterclockwise circuity  ($C=-1$), meaning negative chirality,  was obtained by relaxing the ferromagnetic state in the absence of external magnetic and electric fields.  Then the interaction between the incident electromagnetic field and the magnetic vortex was modeled through an effective magneto-optical coupling, characterized by a coefficient of approximately $10^{-2}$. The interference of two incident optical fields with closely spaced frequencies $\omega_1$ and $\omega_2$ generates an effective magnetic field ($\propto \mathbf{E}(\omega_1) \times \mathbf{E}^*(\omega_2)$) oscillating at the difference frequency $\Omega=\omega_1-\omega_2$, which can lie in the GHz regime (cf. the Supp.\ Mat.). The azimuthal component of this effective field is described by
\begin{equation}
B_{\phi}(\rho,t)
=
B_0 \frac{k_{\perp}}{k_z}
J_0(k_{\perp}\rho) J_1(k_{\perp}\rho)
\exp [i(\omega_1-\omega_2)t] +c.c.,
\end{equation}
where $B_0$ is the effective field-amplitude. Because the two optical frequencies are close, their wave vectors are approximately equal on the characteristic length scale of the vortex, allowing us to adopt $\mathbf{k}_1\simeq\mathbf{k}_2\equiv\mathbf{k}$.  

For a driving frequency of $\Omega = 10$ GHz, the simulations reveal a pronounced sensitivity of the vortex dynamics to the spatial structure of the driving field.  We note that circularly (or linearly) polarized  optical pulses produced an effective homogeneous (or zero) magnetic field (cf. Supp.\ Mat.).  Our numerical simulations evidence that whereas a spatially uniform effective magnetic field  has a negligible effect  under the same  excitation conditions, a structured magnetic field with a characteristic amplitude of approximately $B_0=0.1$ T [cf. Fig. \ref{simulations}(a)] efficiently excites rapid vortex precessional dynamics [cf. Figs. \ref{simulations}(b) and (c)]. These driven magnetization dynamics are subsequently imprinted onto the scattered electromagnetic field, producing spatially structured electric-field patterns with nontrivial topological features [cf. Fig. \ref{simulations}(d)]. Remarkably, an increase of the effective magnetic-field amplitude to approximately 1 T induces a reversal of the vortex-core polarity (cf. Supp.\ Mat.), thus demonstrating the possibility of controlling the magnetic texture through  structured electromagnetic excitations.

The micromagnetic results support the predictions of the analytical RCWM framework and highlight the potential of structured electromagnetic fields for manipulating noncollinear magnetic textures. More broadly, they identify magnetic vortices as a promising platform for reconfigurable electromagnetic-field control, with noncollinear magnetic order providing an additional degree of freedom for tailoring the scattered electromagnetic fields beyond conventional geometry-based approaches. 
We note, intense optical fields need not to be maintained throughout the cycle of magnetization dynamics and switching.  Rectification magnetic  fields due to  optical pulses of some hundred femtosecond duration   can drive  stroboscopically the magnetic  dynamics in a controlled manner \cite{PhysRevLett.102.057204}.  Another possibility is to utilize radially polarized plasmonic fields \cite{Tugchin2015PlasmonicTip,Lerman2009Nanofocusing,Chen2009PlasmonicLens,Zhang2018RadiallyPolarizedBeam,Tugchin2016ConicalPlasmons} that generate localized high-amplitude fields. 
}

\section{Discussion}\label{sec12}

The spherical RCWM developed here provides a framework for treating electromagnetic scattering from spatially nonuniform magnetic textures with anisotropic permeability. By combining a spherical-wave representation with a shell-wise description of the permeability tensor, the method incorporates the vortex's internal magnetic structure directly into the scattering problem. The results show that a magnetic vortex can act as a structured photonic element: its spatially varying magnetization couples incident-field components into complementary polarization channels that are absent from the incident field. These converted fields have characteristic spatial distributions and are accompanied by a nonzero azimuthal component of electromagnetic energy flow, revealing how the magnetic texture is encoded in the scattered electromagnetic field.

The origin of the polarization conversion is the off-diagonal component of the spatially dependent magnetic susceptibility. The response is therefore not simply that of a localized magnetic perturbation; it is shaped by the continuous variation of the local magnetic orientation across the vortex. This result also clarifies that a nonzero skyrmion number is not a prerequisite for texture-mediated polarization conversion. Local magnetic non-collinearity is sufficient to provide spatially varying coupling channels, while global topology may distinguish other classes of textures and their responses.

The comparison between homogeneous and radially polarized illumination demonstrates that the strength of the converted fields depends strongly on the spatial structure of the incident field. Under the conditions considered here, the two illuminations produce qualitatively similar converted-field patterns, but the amplitudes of the converted components differ by one order of magnitude. This comparison highlights the role of spatial overlap between the incident field and the anisotropic susceptibility of the vortex. 

Complementary micromagnetic simulations show polarization-textured optical pulses results in structured effective magnetic field excitation that can drive vortex-core precession and, at sufficiently strong excitation, reverse the core polarity. These simulations illustrate a possible dynamical aspect of the interaction between structured excitation and magnetic textures. They are complementary to, rather than a fully self-consistent continuation of, the spherical electromagnetic-scattering calculation. Taken together, the results motivate further work on linking texture dynamics and optical response within a common multiscale model and on assessing the conditions under which such effects may be realized experimentally. The findings suggest a route toward magnetically reconfigurable photonic elements. In an array of magnetic vortices, elements with identical physical geometry but different circulation or core polarity could produce different scattering responses. The magnetic state could thus provide an additional control parameter for the amplitude, phase, polarization, and angular distribution of scattered fields. The principle emerging from this study is twofold: the magnetic texture determines how the electromagnetic field can be converted, while the spatial structure of the incident field determines how effectively that conversion is excited. This framework provides a basis for exploring magnetic non-collinearity as an active design parameter in structured-light scattering and magnetically reconfigurable photonics.
\section*{Declaration statements}
\bmhead{Data Availability}
Further technical details to reproduce the data are in the Supplementary Information. All details to numerically reproduce the RCWM data are in the main text; these and other data presented in the paper are available from the corresponding author on reasonable request. 

\bmhead{Code Availability}
The codes which have not been further developed during ongoing investigation are available  upon reasonable request.

\bmhead{Acknowledgements}
{
JB thanks the DFG for support under project nr. 429194455.}

\bmhead{Author contributions}

VJ and CLJ carried out the numerical simulations, validation, and testing. All authors contributed to conceptual development, writing, and critical revision of the manuscript.

\bmhead{Competing Interests}

The authors declare no potential competing interests.

\bibliography{sn-bibliography}
\vspace*{1cm}

\appendix
{\textbf{\huge Appendix}}\\

\section{Optical rectification for vector beams and spatially structured excitation}
We are interested in the way rectification fields of radially polarized beams act on the magnetization dynamics. To this end, let us first recall the case of a conventional spatially homogeneous field and introduce the normalized magnetization, which is defined as
\[
\mathbf m(\mathbf r,t)=\frac{\mathbf M(\mathbf r,t)}{M_s},
\qquad |\mathbf m|=1, \] and its dynamics are governed by the Landau--Lifshitz--Gilbert (LLG) equation,
\[ \frac{\partial\mathbf m}{\partial t} =-\frac{\gamma}{1+\alpha^2} \left[ \mathbf m\times\mathbf H_{\mathrm{eff}} +\alpha\mathbf m\times (\mathbf m\times\mathbf H_{\mathrm{eff}}) \right], \]
where $\gamma>0$ is the gyromagnetic ratio and $\alpha$ is the Gilbert damping constant. The effective field includes external, anisotropy, demagnetizing, exchange, and optical contributions:
\[\mathbf H_{\mathrm{eff}} =\mathbf H_{\mathrm{static}}+\mathbf H_{\mathrm{ext}} +\mathbf H_{\mathrm{opt}}, \qquad \mathbf H_{\mathrm{ext}} =\frac{2A}{\mu_0M_s}\nabla^2\mathbf m.\]
For a magneto-optical medium, the inverse Faraday effect (IFE) \cite{vanDerZiel1965,Mills1998}, which is a second-order  in the optical field, produces an effective magnetic field proportional to the optical spin density:
$ \mathbf H_{\mathrm{IFE}} =C_{\mathrm{IFE}}\,i\mathbf E^*\times\mathbf E.$
For circularly polarized light, this reduces to 
$\mathbf H_{\mathrm{IFE}}\propto I\sigma\hat{\mathbf k},$
where $I$ is the optical intensity, $\sigma=\pm1$ is the helicity, and $\hat{\mathbf k}$ is the propagation  (Poynting) direction.
For two optical fields,
$ \mathbf E(t)=\mathbf E_1e^{-i\omega_1t} +\mathbf E_2e^{-i\omega_2t},$ with $ \Delta\omega=\omega_1-\omega_2$,
averaging over the rapid optical oscillations, the IFE field contains individual contributions and a beat-frequency term:
$ \mathbf H_{\mathrm{IFE}}^{\mathrm{av}} =\mathbf H_1+\mathbf H_2+\mathbf H_{\mathrm{beat}}, $
where
$ \mathbf H_j=C_{\mathrm{IFE}}\,i\mathbf E_j^*\times\mathbf E_j $
and
$ \mathbf H_{\mathrm{beat}} =2C_{\mathrm{IFE}}\operatorname{Re} \left[ i\mathbf E_1^*\times\mathbf E_2 e^{i\Delta\omega t} \right]. $
When $\omega_1,\omega_2\gg|\Delta\omega|$, the beat field can drive magnetization at the difference frequency. Efficient excitation occurs when $\Delta\omega$ approaches a magnetic resonance frequency, the case which is studied in the main text.
%
%

\textbf{Structured fields:} For a vector beam, let
$ \mathbf E_j(\mathbf r,t) = A_j(\mathbf r,t) \mathbf e_j(\mathbf r) e^{\ii\phi_j(\mathbf r,t)} $
then we have
$ \ii\mathbf E_1^*\times\mathbf E_2 = \ii A_1^*A_2 \left( \mathbf e_1^* \times \mathbf e_2 \right) e^{\ii(\phi_2-\phi_1)}. $
The beat field is
$           \mathbf H_{\mathrm{beat}}(\mathbf r,t) = 2C_{\mathrm{IFE}}
\operatorname{Re}
\left\{ \ii A_1^*A_2
\left( \mathbf e_1^* \times \mathbf e_2 \right) e^{\ii[ \Delta\omega t+ \phi_2(\mathbf r)- \phi_1(\mathbf r) ]} \right\}.
$
Thus, we identify three  spatially varying contributions due to
\begin{enumerate}
	\item  amplitude profiles $A_1(\mathbf r)$ and $A_2(\mathbf r)$,
	\item  spatially varying polarization vectors
	$\mathbf e_1(\mathbf r)$ and $\mathbf e_2(\mathbf r)$,
	\item and  spatial relative phases
	$\phi_2(\mathbf r)-\phi_1(\mathbf r)$.
\end{enumerate}
Consequently, the magnetic excitation is generally not separable
into a purely temporal function multiplied by a fixed magnetic-field
direction.\\
Let us consider  a paraxial radially polarized beam,
\begin{equation}
	\mathbf e_r(\varphi) = \cos\varphi\,\hat{\mathbf x} + \sin\varphi\,\hat{\mathbf y} = \hat{\mathbf r}.
\end{equation}
{This vector is real. Thus, 
	\begin{equation}
		\mathbf e_r^*\times\mathbf e_r=0.
	\end{equation}
	Therefore, in the  paraxial approximation
	\begin{equation}
		\mathbf H_{\mathrm{IFE}}=0
	\end{equation}
	for a locally linearly polarized radial beam.
	This result does not imply however that a radially polarized beam can never
	produce an optomagnetic response. Under tight focusing, longitudinal
	field components can appear as discussed in the main text. Also, even in the paraxial regime
	two phase-locked  fields with inclined propagation direction may induce an effect. In the non-paraxial regime, generally we have
	\begin{equation}
		\mathbf E = E_r\hat{\mathbf r} + E_z\hat{\mathbf z},
	\end{equation}
	in which case the local polarization becomes elliptical, and
	\begin{equation}
		\ii \mathbf E^*\times\mathbf E\neq0.
	\end{equation}
	Therefore,  a spatially structured IFE field can arise.
	Note, when the optical field varies spatially, the exchange contribution
	cannot generally be neglected.
	The effective field is
	$	\Heff = \mathbf H_0 + \mathbf H_{\mathrm{ani}} + \mathbf H_{\mathrm{demag}} + \frac{2A}{\mu_0M_s}\nabla^2\m  \mathbf H_{\mathrm{IFE}}(\mathbf r,t).$
	Noting that  the  spatial Fourier components of the optical field  are
	$ \mathbf H_{\mathrm{IFE}}(\mathbf r,t) = \int \frac{d^3k}{(2\pi)^3} \, \mathbf H_{\mathrm{IFE}}(\mathbf k,t) e^{\ii\mathbf k\cdot\mathbf r}, $
	we conclude that a vector beam can excite a distribution of magnetic wavevectors.
	In a translationally invariant approximation, each Fourier component
	can be treated separately:
	$	\m(\mathbf r,t) = \sum_{\mathbf k} \m_{\mathbf k}(t) e^{\ii\mathbf k\cdot\mathbf r}.$
	The exchange field then becomes $	\mathbf H_{\mathrm{ext},\mathbf k} = - \frac{2A}{\mu_0M_s} k^2 \m_{\mathbf k}.$
	Consequently, the spatial structure of the optical beam determines
	which magnetic spin-wave wave vectors are driven.
	
	\section{Convergence test for the spherical rigorous coupled wave method}
	
	\begin{figure}[t]
		\centering
		\includegraphics[width=0.8\linewidth]{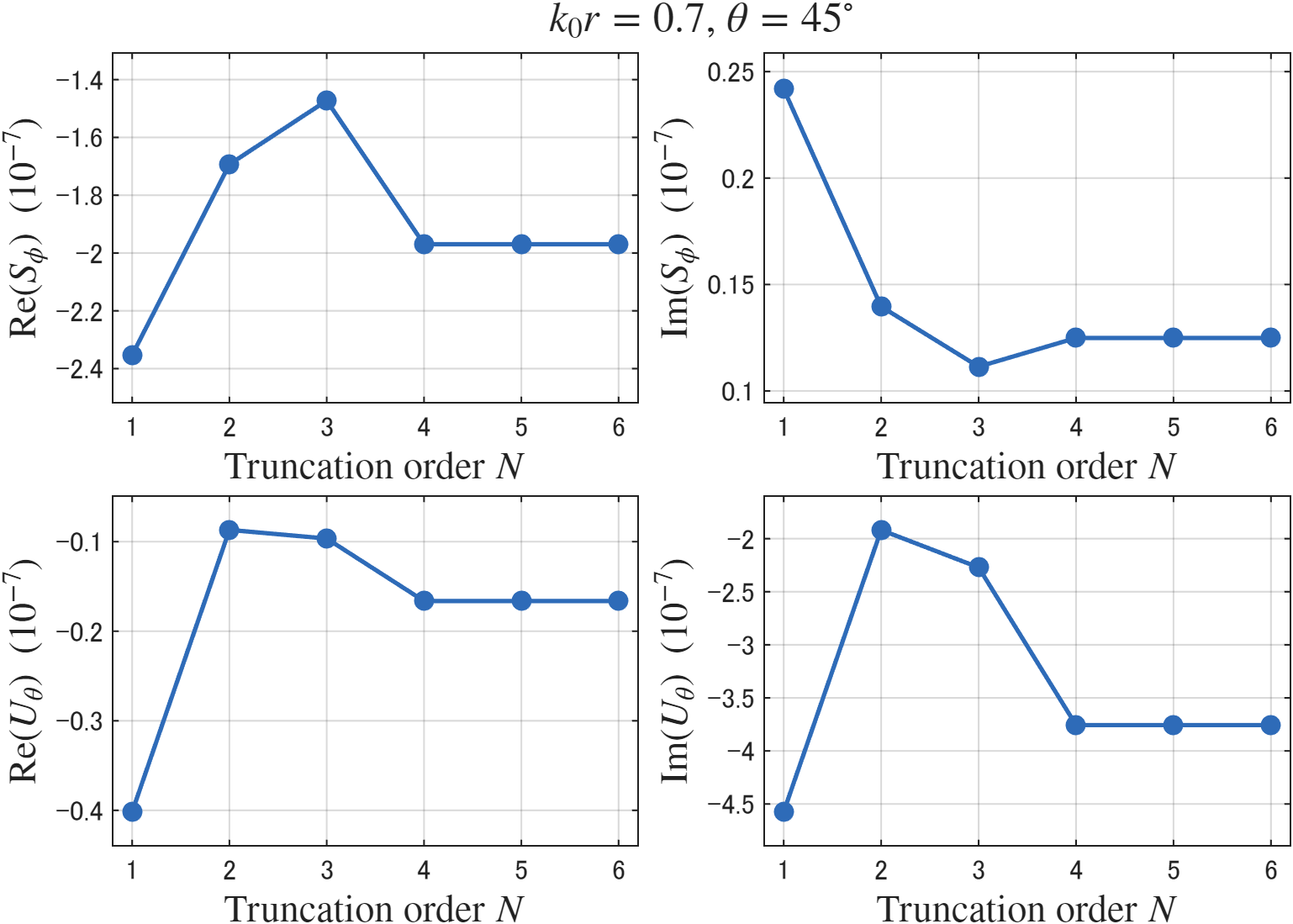}
		\caption{
			Convergence of the electromagnetic field components with increasing Fourier truncation order $N$. The upper panels show the real and imaginary parts of $S_{\phi}$, while the lower panels show the corresponding components of $U_{\theta}$. The calculations are performed at the fixed observation
			point $k_0r=0.7$ and $\theta=45^\circ$.}
		\label{convergence}
	\end{figure}
	
	{To examine the numerical convergence of the RCWM formulation, we investigate the dependence of the field components $S_{\phi}$ and $U_{\theta}$ on the Fourier truncation order $N$. The calculations are performed at a fixed observation point, $k_0r=0.7$ and $\theta=45^\circ$, while successively increasing the number of Fourier space harmonics. Figure~\ref{convergence} presents the real and imaginary parts of both field components, $S_{\phi}$ and $U_{\theta}$. Although noticeable variations are observed for relatively small truncation orders, the calculated field amplitudes stabilize as $N$ increases. In particular, the results exhibit negligible variations for $N\geq4$, indicating convergence at the selected observation point. It should be noted, however, that increasing the truncation order does not necessarily lead to improved numerical accuracy. The inclusion of higher Fourier space harmonics increases the dimensions of the RCWM matrices and may amplify numerical errors associated with the conditioning of the resulting linear system of equations. Consequently, a moderate truncation order is adopted to achieve a reasonable balance between the accuracy of the Fourier representation and the numerical stability of the solution.}
	
	\section{Micromagnetic simulation at 1 T}\label{sec3}
	In the main text we discussed the action of polarization structured fields on the magnetization dynamics in the linear response regime. The simulations presented in Fig.(\ref{magnetic})  demonstrate the effectiveness  of using optical polarization texturing for driving strongly the magnetic dynamics and even flipping the polarity and the chirality of the vortex within nanoseconds.
	\begin{figure}[t]
		\centering
		\includegraphics[width=0.9\linewidth]{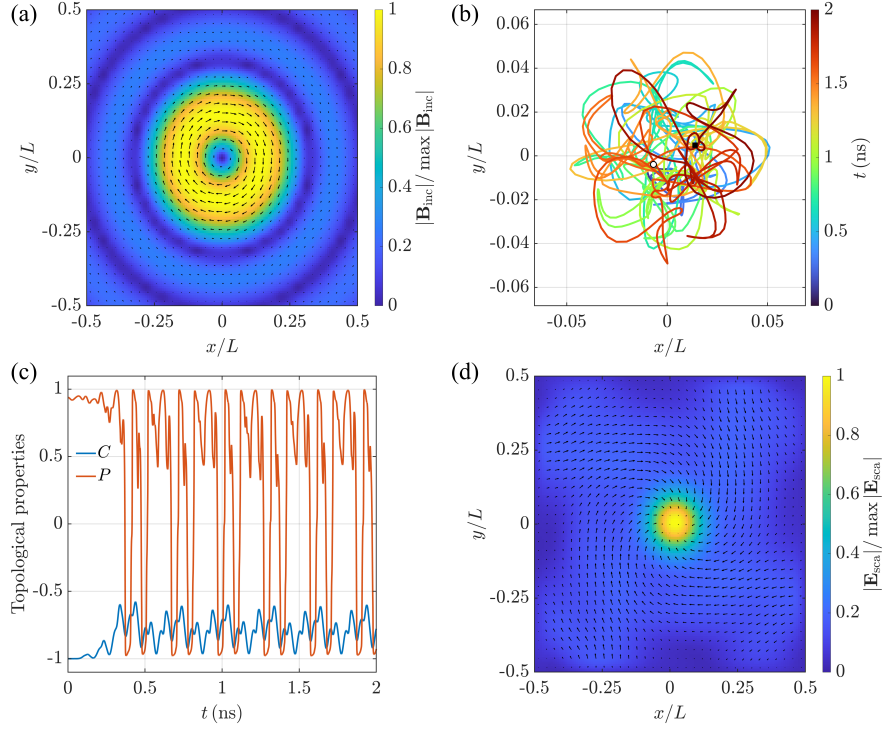}
		\caption{
			Micromagnetic simulations of magnetic vortex dynamics and
			electromagnetic-field scattering. (a) Spatial distribution of the effective magnetic field with an amplitude of 1 T, generated by the interference
			of two structured optical electric fields. (b) Trajectory of the magnetic vortex core under the structured-field excitation. (c) Time evolution of the vortex-core polarity ($P$) and circuity ($C$) under excitation by the effective
			magnetic field shown in (a). The figure demonstrates chirality $\chi=C\, P$ switching on ns time scale. (d) Snapshots of the spatial distribution of the emergent electric field induced by the driven vortex dynamics.}
		\label{magnetic}
	\end{figure}

\end{document}